
\magnification=1200
\parindent 1.0truecm
\baselineskip=16pt
\rm
\null


\footline={\hfil}
\rightline{hep-ph/9804344}
\rightline{\bf DFUPG--98--GEN--02}
\rightline{\sl April 1998}
\vglue 2.0truecm

\centerline{\bf $KN$ SIGMA TERMS, STRANGENESS IN THE NUCLEON, }
\centerline{\bf AND DA$\Phi$NE$^{\dag}$ }
\vglue 1.0truecm

\centerline{Paolo M. Gensini}
\centerline{\sl Dip. di Fisica, Universit\`a, and Sez. I.N.F.N., Perugia, 
Italy}
\vglue 4.0truecm

\centerline{Talk presented at the}
\centerline{{\bf ``1998 LNF Spring School''},}
\centerline{\sl Frascati, 14 -- 18 April 1998.}
\centerline{Expanded and updated version of a talk originally presented at}
\centerline{\bf ``IVth Int. Symp. on Pion--Nucleon Physics }
\centerline{{\bf and the Structure of the Nucleon''},}
\centerline{\sl Bad Honnef, 9 - 13 September 1991,}
\centerline{published in {\sl $\pi N$ Newletter} {\bf 6} (1992) 21--32.}
\vglue 3.0truecm

\hrule
\vglue 0.5truecm
\item{\dag)} Research supported by the E.E.C. Human Capital and Mobility 
Program under contract No. CHRX--CT92--0026.

\pageno=0
\vfill
\eject

{\ \ \ \ \ \ }
\pageno=0
\vfill
\eject

\footline={\hss\tenrm\folio\hss}
\vglue 0.5truecm

\centerline{\bf $K N$ SIGMA TERMS, STRANGENESS IN THE NUCLEON, }
\centerline{\bf AND DA$\Phi$NE }
\vglue 0.3truecm

\centerline{Paolo M. Gensini}
\centerline{\sl Dip. di Fisica, Universit\`a, and Sez. I.N.F.N., Perugia, 
Italy}
\vglue 2.0truecm

\leftline{\bf 1. INTRODUCTION.}
\vglue 0.3truecm

This talk will consist of three parts: in the first I shall stress 
the importance of the $K N$ sigma terms in fixing the composition of the 
baryonic ground states, and particularly the scalar, strange--quark density 
in the nucleon. I shall also briefly cover some of the reasons, besides 
QCD, why it is important to know with some accuracy such 
strong--interaction parameters, and delve on why they can not simply be 
inferred from the (already rather well known) $\pi N$ one.

In the second chapter, I shall review the situation of their extraction 
from data, using as a guideline the clearer $\pi N$ case, and discuss 
merits and shortcomings of three different methods. I shall try to cover 
the last two decades of studies in the field, i.e. those starting with 
A.D. Martin's analysis of low-energy $K N$ systems$^1$, and show that, 
though there is still no consensus, the most reliable methods indicate 
a large value for the isoscalar parts of the $K N$ sigma terms, and can 
only put rather generous bounds on their isovector parts.

A third, final section will be dedicated to experimental outlooks on 
the future of low-energy $K N$ physics, focussing on the possibilities that 
are opening up at $\phi$--factories: due to their high design luminosities 
and to the kaon production mechanisms, these will be {\sl almost 
monochromatic sources of extremely--low--background, low--momentum kaons.} 
The challenge is how to exploit these kaons for scattering experiments in 
a geometry radically different from those we have been accustomed up to 
now. Not being a rugged experimentalist, I shall limit myself to a 
conceptual sketch of a {\sl dedicated} detector and to some 
``back--of--the--envelope'' calculations, which I hope will show that 
high--statistics measurements should indeed be feasible with advanced, 
currently available technologies. I shall also briefly list the 
relevant measurements already possible at the existing three DA$\Phi$NE 
experiments DEAR, FINUDA and KLOE.
\vfill
\eject

\noindent{\bf 2. $\sigma$--TERMS AND ``MEASUREMENTS'' OF THE STRANGE--QUARK 
SCALAR DENSITY OF THE NUCLEON.}
\vglue 0.3truecm

The problems, posed by the $\pi N$ $\sigma$--term being larger than 
expected on the basis of the simplest quark--model pictures of the 
nucleon, have been with us for quite a while$^2$, before Donoghue and 
Nappi$^3$ pointed to this fact as to an indicator of a large $\bar ss$ 
component in the nucleon sea, foreign to the then standard 
quark--model pictures, but not unexpected in a Skyrmion picture of the 
nucleon$^4$.

However, only a few authors have stressed$^5$ that the $\pi N$ 
$\sigma$--term is {\sl not} the best indicator of a scalar strange--quark 
density in the nucleon sea, but {\sl just} the quantity sensitive to the 
latter that, as of today, we know the best.

Indeed, the evidence it provides is quite indirect, resting on {\sl two} 
other assumptions about the {\sl precise} values of {\sl both} the 
quark--mass ratio $2m_s/(m_u+m_d)$ {\sl and} the $SU(3)_f$--breaking 
terms in the octet--baryon masses. What is ``measured'' is indeed the 
proton expectation value of the operator
$$
\sigma_{\pi^+\pi^-}(x)={1\over2}\ (m_u+m_d)\ [\bar u(x)u(x)+\bar d(x)d(x)]
\ ,\eqno(1)
$$
which clearly {\sl does not} gauge directly the strange--quark density 
$\bar s(x)s(x)$. The above operator is a pure isoscalar: the corresponding 
operator for {\sl neutral} pions 
$$
\sigma_{\pi^0\pi^0}(x)=m_u\ \bar u(x)u(x)+m_d\ \bar d(x)d(x)\eqno(1')
$$
has a small, additional isovector part coming from the $SU(2)_f$--violating 
part of the Hamiltonian, and related via $SU(2)_f$ to the soft--pion limit 
of the charge--exchange, {\sl crossing--even} amplitude. This part is 
expected to be suppressed by at least one order of magnitude with respect 
to the former, {\sl even when} exerting the caution suggested by the 
recent observation of sizeable departures from the Gottfried sum rule in 
deep inelastic scattering$^6$.

For sake of brevity I shall neglect here these $SU(2)_f$--violating 
effects, being their expected contibutions well below present experimental 
(and theoretical) capabilities. To gauge {\sl directly} the scalar density 
$\bar s(x)s(x)$ one should turn instead to operators like
$$
\sigma_{K^+K^-}(x)={1\over2}\ (m_s+m_u)\ [\bar u(x)u(x)+\bar s(x)s(x)]
\eqno(2)
$$
(the analogous operator for $K^0$'s can be obtained replacing $u$'s by 
$d$'s in Eqn. 2), and to the corresponding one for the octet component of 
the $\eta$--meson 
$$
\sigma_{\eta_8\eta_8}(x)={1\over3}\ \sigma_{\pi^0\pi^0}(x)+{4\over3}\ m_s 
\bar s(x)s(x)\ ,\eqno(3)
$$
which, even if not directly measurable, still plays an important role in 
meson--condensation phenomena in dense nuclear matter$^7$ and is {\sl 
directly} related to the chiral--symmetry breaking part of the 
standard--model Hamiltonian
$$
H_{SB}(x)={3\over4}\ [\sigma_{\eta_8\eta_8}(x)+\sigma_{\pi^0\pi^0}(x)]\ ,
\eqno(4)
$$
responsible for the shift of the nucleon mass from its chiral--symmetry 
value (at lowest--order in the symmetry breaking),
$$
\Delta M_N^{(0)}\simeq\langle N\vert H_{SB}(0)_{I=0}\vert N\rangle=
{3\over4}\ [\Sigma_{\pi^\pm N}+\Sigma_{\eta_8N}]\ . \eqno (5)
$$
Last but not least, we also wish to mention here the dominantly isovector 
operator $\sigma_{\pi^0\eta_8}(x)$, of some relevance in the study of the 
$I=1$ $\bar{K}N$ $t$--channel amplitudes, given as
$$
\sigma_{\pi^0\eta_8}(x)={1\over{\sqrt3}}[m_u\bar u(x)u(x)-m_d\bar d(x)d(x)]
\ . \eqno (5')
$$

Separating $H_{SB}(x)$ into its singlet and both isoscalar and isovector 
octet components respectively as 
$$
H_{SB}(x)=H_{SB}^{(0)}(x)+H_{SB}^{(8)}(x)+H_{SB}^{(3)}(x)\ , \eqno (6)
$$
where 
$$
H_{SB}^{(0)}=m_0S_0={1\over3}\ (m_u+m_d+m_s)(\bar uu+\bar dd+\bar ss)\ , 
\eqno (6')
$$
$$
H_{SB}^{(8)}=m_8S_8={1\over6}\ (m_u+m_d-2m_s)(\bar uu+\bar dd-2\bar ss)\ ,
\eqno (6'')
$$
and 
$$
H_{SB}^{(3)}=m_3S_3={1\over2}\ (m_u-m_d)(\bar uu-\bar dd)\ , \eqno (6''')
$$
we can express the meson--nucleon $\sigma$--terms (considering only 
``elastic'' channels, and charged pions and kaons) as
$$
\Sigma_{\pi^{\pm}N}={m\over{m_s-m}}\ M_8 \ {1\over{1-y}}\ , \eqno (7)
$$
$$
\Sigma_{K^{\pm}N}^{(0)}={{m_s+m_u}\over{m_s-m}}\ {1\over4}\ M_8\ {{1+y}
\over{1-y}}\ , \eqno (8)
$$
and
$$
\Sigma_{\eta_8N}={m_s\over{m_s-m}}\ {2\over3}\ M_8\ {{y+m/(2m_s)}\over
{1-y}}\ , \eqno (9)
$$
where we have neglected small, $SU(2)_f$--violating terms from the nucleon 
wave function, introduced the ``nuclear isospin'' notation $\Sigma^{(I=0,1)}=
{1/2}\ [\Sigma_p+(-1)^I\Sigma_n]$ (useful to work in nuclear matter) plus 
Gasser's notation$^8$ $y=2\langle N\vert\bar ss\vert N\rangle/\langle N\vert
(\bar uu+\bar dd)\vert N\rangle$, and defined
$$
M_8=-{1\over2}\int d^3\vec x\ [\langle p\vert\lbrace3\ H_{SB}^{(8)}(\vec x,0)
\rbrace\vert p\rangle + (p\to n)]\ . \eqno (10)
$$
The scale of ``isovector'' combinations (and $SU(2)_f$--violating terms) is 
set instead by
$$
M_3=-{1\over2}\int d^3\vec x\ [\langle p\vert\lbrace\sqrt3\ {m_8\over m_3}\ 
H_{SB}^{(3)}(\vec x,0)\rbrace\vert p\rangle - (p\to n)]\ , \eqno (11)
$$
which we shall not use extensively, but which has however to be kept under 
close scrutiny to ensure the absence of ``large'' $SU(2)_f$--violating terms 
in the soft--meson limit. These ``isovector'' parts can be easily written 
down using the above notation$^9$, but will not be considered here as they 
are independent on $y$; they are not negligible, at least for kaons, and 
influence detailed analyses: for instance their neglect in $K$--condensation 
calculations$^7$ has masked till now an interesting consequence for 
supernov\ae$^{10}$ (probably already seen in the IMB and Kamiokande 
neutrino signals from Shelton's supernova, SN 1987A$^{11}$), {\sl i.e.} 
the possible presence of an ``energetic'', pure $\nu_{\mu}$ signal a short 
time after the ``thermal'' neutrino burst from the gravitational collapse.

Putting together eqs. (5), (7) and (9), one obtains 
$$
\Delta M_N^{(0)}\simeq{m_s\over{m_s-m}}\ M_8\ {{m/m_s+y/2}\over{1-y}}=
\Sigma_{\pi^{\pm}N}\ \big(1+{m_s\over{2m}}\ y\big)\ , \eqno (12)
$$
so that even a value of $y$ as small as 0.2 can make $\Delta M_{SB}$ 
quite larger than $\Sigma_{\pi^\pm N}$, the traditional, quark--model 
expectation for the chiral--symmetry--breaking shift in the nucleon 
mass.

The two mass scales (10) and (11) were traditionally calculated from 
octet--baryon masses at lowest order in the symmetry breaking: they can 
however receive non--negligible corrections from higher--order terms. 
Already Gasser$^8$ found a sizeable correction to $M_8$, working at one 
loop in chiral perturbation theory: we expect that going to higher orders, 
or higher number of loops, could increase $M_8$ even further (see the recent 
re--evaluation of the ``scalar'' pion form factor$^{12}$, yielding a very 
``soft'' result, in line with our dispersive estimate of eighteen years 
ago$^{13}$).

Note that higher orders in the symmetry breaking (with $m_u\neq m_d$) 
break also the isospin invariance of the nucleon wave function, so that 
$\langle p\vert H_{SB}^{(8)}\vert p\rangle\neq\langle n\vert H_{SB}^{(8)}
\vert n\rangle$ and $\langle p\vert H_{SB}^{(3)}\vert p\rangle\neq-\langle 
n\vert H_{SB}^{(3)}\vert n\rangle$: however the size of the discrepancy 
is of $O(m_3/m_8\simeq 2\cdot10^{-2})$ with respect to the 
$SU(2)_f$--symmetric values, and thus not as important as the rest of the 
contribution.

To try and estimate these higher--order effects, {\sl independently} 
of either the Bern group approach, or Skyrmion phenomenologies of all, 
different kinds, I have taken the rather na\"\i ve approach of working 
in a Hamiltonian formalism, and used second--order Raleigh--Schr\"odinger 
perturbation theory. Restricting the mixing of the baryon octet to just 
{\sl one} representation for each {\sl non--exotic} multiplicity, I have 
found$^{9,14}$ for the two mass--breaking scales
$$
M_8=[626\ \hbox{MeV}]+[(200\pm20)\ \hbox{MeV}+8\cdot\Sigma] \eqno (13)
$$
and
$$
M_3=[132\ \hbox{MeV}]-[(35\pm6)\ \hbox{MeV}]\ , \eqno (13')
$$
where in each expression the first and second square bracket represents, 
respectively, the first-- and second--order flavour--symmetry--breaking 
contribution, and $\Sigma\geq0$ is the unitary--singlet--admixture term 
in the mass of the $\Lambda$--hyperon. The latter can not vanish if we 
are to reproduce, {\sl in the same formalism}, flavour--symmetry--breaking 
effects in the axial--vector couplings$^{14,15}$, and is better to be 
strongly limited from above if the octet has to stay lighter than the 
decuplet in the symmetry limit: one can thus estimate $M_8$ to lie between 
a minimum of about 850 MeV and a maximum which cannot exceed 1,150 MeV, or 
$M_8\simeq(1,000\pm150)$ MeV, somewhat above Gasser's one--loop 
estimate$^8$, which can be translated in our language into the value 
$M_8=(840\pm120)$ MeV.

Note that to extract $y$ from eq. (7) one would also have to know the 
strange--to--non--strange quark mass ratio $2m_s/(m_u+m_d)$, for which 
Gasser used (consistently) the one--loop result $m_s/m\simeq25$. However, 
QCD sum rules$^{16}$ give a wider range of values for this ratio, so as to 
make its precise value questionable: a careful assessment of all the 
uncertainties makes a value of $y \simeq 0$ {\sl not incompatible} with 
the value $\Sigma_{\pi^\pm N}\simeq50$ MeV, on which consensus seem to have 
been finally reached among the different methods$^{17,18}$, if all 
theoretical uncertainties {\sl both} on $M_8$ {\sl and} on $m_s/m$ are 
pushed toward their upper limits.

By inspecting eq. (8) one can see that: i) $\Sigma_{K^{\pm}N}^{(0)}$ is 
very little dependent on $m_s/m$ for not too small values of the ratio, 
and ii) much more dependent on $y$ than $\Sigma_{\pi^\pm N}$. The sad 
note is that, despite all efforts including mine, we are far from 
reaching consensus but for its order of magnitude, 
expected to be of several hundred MeV's. Since this lack of consensus is 
due in part to the theoretical difficulties inherent to an extrapolation 
over much larger four--momentum intervals than in the $\pi N$ case, and in 
part to the poorer information coming from experiments on low--energy 
$\bar KN$ systems, we shall devote the following two sections first to a 
review of the extrapolation methods, and then to an outlook on possibilities 
opening up at the DA$\Phi$NE $\phi$--factory.

\vglue 0.6truecm
\noindent{\bf 3. METHODS OF EXTRAPOLATION TO $q^2 = t = \omega^2 = 0$: A 
SUBJECTIVE REVIEW.}
\vglue 0.3truecm

At the first presentation of such a summary, in 1991 at Bad Honnef$^{19}$, I 
updated the (unpublished) report presented in 1982 at the Black Forest 
Meeting in Todtnaueberg, touching only passingly results and methods where no 
improvements had been registered, and concentrating instead on those 
which had been improved upon after that date. Here the main improvement 
over Bad Honnef will be a revision of the ``scalar form factors'' 
following their more recent theoretical re--evaluations and the 
re--analyses of low--energy $\pi\pi$ data prompted by the activities 
of the DA$\Phi$NE Theory Group$^{20}$.

The matrix elements of the operators discussed in the previous section, 
generally known as the $\sigma$--terms, are better to be seen (from 
$su(3)\times su(3)$ current algebra and PCAC) as the 
zero--energy, zero--momentum--transfer values of the scattering amplitudes 
for massless mesons, shorn of their eventual pseudovector--coupling Born 
terms. For a process $a + B \to b + B'$ (where the mesons $a$ and $b$ are 
to be taken off their mass shells), the kinematics are defined by the 
variables $\vec{P} = \{q_a^2, q_b^2, \omega, t \}$, where $\omega = 
(s-u)/2(M+M')$, since energy--momentum conservation fixes $s+u = M^2 + M'^2 
+ q_a^2 + q_b^2 - t$, and the $\sigma$--term is thus defined as
$$
\Sigma_{aB\to{bB'}}=\langle{B'}\vert\sigma_{a\bar{b}}(0)\vert{B}\rangle=
-\hbox{lim}_{\vec{P}\to\vec{O}}\ {{f_af_b}\over2}\big[A_{aB\to{bB'}}
(\vec{P})-A^{Born(pv)}_{aB\to{bB'}}(\vec{P})\big]\ ,
\eqno (14)
$$
where $\vec{O} = \{0, 0, 0, 0\}$.

At least three methods have been widely used in the literature (I choose 
deliberately not to mention those less recommendable or of dubious 
validity): i) the ``improved'' Altarelli--Cabibbo--Maiani 
technique$^{21,22}$, ii) ``modified'' Fubini--Furlan sum rules applied 
to $K^-$--nucleus scattering lengths$^{23,24}$, and iii) a ``unitarized'' 
version of the Cheng--Dashen relation$^{13,25}$.

\vglue 0.3truecm
\noindent{\bf 3.1. THE ``IMPROVED'' ALTARELLI--CABIBBO--MAIANI METHOD. }
\vglue 0.3truecm

The method originally devised by Altarelli, Cabibbo and Maiani$^{26}$ 
to extract the $\pi N$ $\sigma$--term, and subsequently extended to the 
$KN$ ones by Reya$^{21}$ and by Violini and coworkers$^{22}$, consists 
essentially in continuing, from the threshold to $q^2 = t = \omega^2 = 0$, 
the elastic, crossing--even amplitudes from which all low--mass, 
pseudovector--coupling pole terms have been explicitly subtracted, 
considering such a difference to be adequately described by a truncated 
power series of the above invariants. The original, essential 
shortcomings of the seminal papers$^{21,26,27}$ were soon corrected 
using, instead of the amplitudes at threshold (of course a singular 
point), the zero--energy amplitudes derived from fixed--$t$ dispersion 
relations$^{22,28}$.

The quality of such an approach for the $KN$ systems can be gauged by 
the estimates $\Sigma_{K^\pm p} = 175 \pm 890$ {\sl (sic)} MeV and 
$\Sigma_{K^\pm n} = 718 \pm 460$ MeV reported by G. Violini and his 
coworkers$^{22}$: they also give, for the ``isoscalar'' part 
$\Sigma_{K^\pm N}^{(0)}$ the value $599 \pm 374$ MeV: error estimates 
are thus of the same order or even larger than the $\sigma$--terms 
themselves. The same authors point out that the method {\sl requires} 
large cancellations between terms each one of which, though correlated 
to the others, carries a large uncertainty, mainly due to the poor 
quality of what where (and {\sl still are}) the {\sl best available} 
low--energy $\bar{K}N$ data.

Note that the same method applied to $\pi N$ amplitudes gave$^{27,28}$ 
the result (rounding figures and giving a {\sl personal} re--evaluation 
of the {\sl original} errors) $\Sigma_{\pi^\pm N} \simeq 50 \pm 10$ 
MeV, not far from modern estimates coming from different methods$^{17,18}$: 
in this case the difference, apart from the higher quality of the data, 
reached already in the late seventies, is to be attributed mostly to the 
analytic structure of the low--energy $\bar K N$ unphysical region, 
responsible of the huge cancellations present in the $K N$ case, and 
totally absent in the $\pi N$ one.

\vglue 0.3truecm
\noindent{\bf 3.2. THE FUBINI--FURLAN ANALYSIS OF MESON-NUCLEUS SCATTERING 
LENGTHS. }
\vglue 0.3truecm

The second method to be briefly reviewed here has been introduced by 
this author a couple of decades ago$^{29}$, and later re-considered$^{17,24}$ 
for application to the world set of mesonic--atom data. It was originally 
motivated by the observations that current--algebra sum rules, derived in 
the collinear frame by Fubini and coworkers$^{30}$ (relating the 
$\sigma$--terms to the amplitudes at threshold), take a much simpler form 
if one can send the target mass to infinity, and that the extreme 
non--smoothness of the low--energy, meson--nucleus scattering amplitudes, 
due to nuclear excitations, can be easily eliminated, summing these 
excitations with standard nuclear sum--rule techniques$^{29}$. Furthermore, 
divergences (appearing in QCD from integrations up to infinite energy 
and virtuality) can be avoided by using a finite--contour version of the 
sum rules, owing to the large mass gap present in the pseudoscalar--meson 
mass spectra.

For $\Sigma_{\pi^\pm N}$ such a method produces an estimate of about 
$48 \pm 9$ MeV, reproducing nicely all detailed features of the 
data$^{17}$ (down to typically nuclear, shell--structure effects), 
available for {\sl separated} isotopes up to $^{27}$Al, plus an 
extrapolation to threshold from a phase--shift analysis of $\pi^\pm - 
^{40}$Ca elastic scattering.

For kaonic atoms one can use data up to uranium (due to the dominantly 
S--wave nature of the interaction), but generally these are available for 
{\sl natural} isotopic mixtures only, so that both isotopic--spin 
dependence and shell-structure effects can not be separated out. Depending 
on assumptions on the renormalization of the hyperon axial couplings in 
nuclear matter, one estimates $\Sigma_{K^\pm N}^{(0)}$ to range from 480 
to 650 MeV, with {\sl purely statistical} errors from the fits$^{23}$ of 
the order of 20 to 30 MeV.

The comparison with the $\pi N$ case shows clearly that 
this analysis is limited by its systematics, which can not be resolved 
(as done in the $\pi N$ case) as long as we can not use data from 
isotopically separated atomic species; thus the different effects are 
lumped together in a global fit to the mass--number dependence, which 
gives too large a weight to 
the heaviest--atom data, precisely those for which the optical--potential 
model used to extract the kaon--nucleus scattering lengths is more open 
to questioning$^{31}$.

\vglue 0.3truecm
\noindent{\bf 3.3. THE ``UNITARIZED'' VERSION OF THE CHENG--DASHEN THEOREM. }
\vglue 0.3truecm

The third approach is an improvement over the rather oversimplified, 
linear expansion originally employed by Cheng and Dashen for the $\pi N$ 
amplitude$^{32}$, and improperly extended to $K N$ ones by some 
authors$^{33}$; however, the original idea can be correctly rephrased 
by stating that all {\sl pseudovector} Born terms of the spin--averaged 
scattering amplitudes vanish exactly along the line $\Gamma_{CD}$, 
defined by $q_1^2/m_1^2 = q_2^2/m_2^2$, $\omega^2 = 
0$, $t = q_1^2 + q_2^2$, so that this line can be used as an 
extrapolation path to go from the current algebra point $q_1^2 = q_2^2 
= \omega^2 = t = 0$ to the mass--shell point $\omega^2 = 0$, $t = m_1^2 + 
m_2^2$.

The major contributions to the amplitude curvature along this line 
are of course expected from the low--mass portion of the $t$--channel 
cuts$^{13}$, while minor contributions are also expected from the 
discontinuities in the mass variables $q_i^2$. The improvement to the 
extrapolation comes from the further observation that, {\sl if}~~the 
discontinuities are dominated by the S--waves and {\sl if}~~one can write 
these latter in an $N/D$ decomposition, Watson's theorem holds both on and 
off the mass shell, and one has$^{13,25}$
$$
A_{\pi^\pm N}^+(2m_{\pi}^2)\ \simeq\ {{2 \Sigma_{\pi^\pm N}}\over
{f_{\pi}(0)^2}}\ \Phi_{\pi\pi}(2m_{\pi}^2)\ ({{m_{\pi\prime}^2}\over
{m_{\pi\prime}^2-m_{\pi}^2}})^2\ ,
\eqno (15)
$$
and 
$$
A_{K^\pm N}^+(2m_K^2)_I\ \simeq\ {{2\Sigma_{K^\pm N}^{(I)}}\over{f_K(0)^2}}\ 
\Phi_{\bar KK}^{(I)}(2m_K^2)\ ({{m_{K\prime}^2}\over{m_{K\prime}^2-m_K^2}})^2\ 
, \eqno (15')
$$
where the Omn\`es function in the second case is related to the first one 
by the $N/D$ decomposition as
$$
\Phi_{\bar KK}^{(I=0)}(t) \ = \ 1 + R \cdot [\Phi_{\pi\pi}(t) - 1]
\eqno (16)
$$
(and $\Phi_{\bar KK}^{(I=1)}(t) \simeq \Phi_{\pi\eta}(t)$, since $R \simeq 1$ 
in the latter case), where 
$$
R = {{m_K^2}\over{{\sqrt6}m_\pi^2}}\cdot{\Sigma_{\pi^\pm N}\over
\Sigma_{K^\pm N}^{(0)}}\ , \eqno (17)
$$
and the {\sl same} Omn\`es functions can be used {\sl on} and {\sl 
off}~~the mass shell. These Omn\`es functions ({\sl a.k.a.} ``the pion 
scalar form factor'' in the $\pi N$ case$^{18}$) have obviously in 
the variable $t$ all the analytical properties of a form factor {\sl on 
the line} $\Gamma_{CD}$: the criticism raised by Coon and 
Scadron$^{34}$ is just a semantic misunderstanding.

It might appear, from the last relation, that $\Sigma_{K^\pm N}^{(0)}$ 
can not be extracted from the on--shell amplitudes, since $\Phi_{\bar 
KK}^{(0)}$ depends on it: but one can, using $R$ as a parameter, derive 
$\Sigma_{K^\pm N}^{(0)}$ from a consistency condition, since its 
dependences on $R$ coming from the two relations (15) and (17) are 
remarkably different$^{25}$. Of course, one has to rely on a simultaneous 
determination of $\Sigma_{\pi^\pm N}$, {\sl possibly} within the same method 
for internal consistency. Using the values calculated by Oades$^{35}$ for 
the zero--energy, pole--term--subtracted, non--flip amplitudes $D_{K^\pm 
p}^+$ and $D_{K^\pm n}^+$ at different values of $t \leq 0$, we 
reconstructed, adding the proper hyperon poles, the two zero--energy, 
spin--averaged amplitudes $A^+_I(t)$ for $I =$ 0, 1. At this point these 
latter were divided by the Omn\`es functions $\Phi_{\bar KK}^{(I)}(t)$ and, 
subtracting again the pseudovector hyperon Born terms (which vanish at the 
Cheng--Dashen point), we obtained two functions$^{25}$ which extrapolated 
smoothly to $t=2m_K^2$, provided we used a conformal mapping of the 
complex $t$--plane to ensure stability, taking there the values 
$[2\Sigma_{K^\pm N}^{(I)}/f_K(0)^2]\cdot[m_{K\prime}^2/(m_{K\prime}^2-
m_K^2)]^2$.

The method, using as inputs the $\pi\pi$ S--waves$^{36}$ and the 
pseudoscalar excitations' masses from recent compilations$^{37}$, gives 
$\Sigma_{\pi^\pm N} \simeq 50$ MeV and $\Sigma_{K^\pm N}^{(0)} \simeq 460$ 
MeV, with the ``physical'' meson decay constants $f_\pi = 132$ MeV and 
$f_K = 154$ MeV. Errors are difficult to assess in this case: for the $\pi 
N$ system such a unitarization method should do no worse than chiral 
perturbation theory (and has indeed been successfully checked against 
Gasser's scalar form factor$^{12,18}$); for the $K N$ systems even 
the dispersion relation results for the zero--energy amplitudes supplied 
by Oades $^{35}$ did not carry any error estimate to start with: varying 
the coupling constants between the extremes used in the dispersion 
relations gives the extrapolation a purely systematic uncertainty of the 
order of 95 MeV, already larger than that associated to the uncertainties 
in the scalar form factor evaluation.

For the $I = 1$ channel, describing $\pi\eta$ and $\bar KK$ couplings 
to the scalar $a_0$--meson by a K--matrix with the observed mass and 
width of the resonance$^{37}$, one can extract by the same technique 
$\Sigma^{(1)}_{K^\pm N} \simeq 78^{+36}_{-56}$ MeV (where again the 
errors are only measures of the dependence of the extrapolation on 
the coupling constants), to be compared with an expectation (to second 
order in $SU(3)_f$--breaking) of $\simeq 25$ MeV.

The method is stable, at least with respect to the 
kaon--nucleon--hyperon couplings: the above uncertainty cover also the 
cases in which, {\sl e.g.} $g_{K\Sigma N}^2$ was put equal to zero; the 
same can not be said for the first of the three methods, derived from the 
Altarelli--Cabibbo--Maiani technique$^{22}$. It is also non--perturbative, 
and general enough to be tested in the $\pi N$ system as well, where the 
perturbative techniques seem not in contradiction with its 
results$^{12,18}$. We are therefore waiting only too eagerly for new 
$\bar KN$ data to put it to even more stringent tests$^{38}$.

A further comment is in order on the previous presentation: all methods 
resting on experimental information from the $\bar KN$ amplitudes are 
presently suffering from the {\sl extremely poor} quality of our knowledge 
of the $S=-1$ meson--baryon systems at low energy, even on its most 
fundamental parameters such as the PBB coupling constants. {\sl All 
information} on the $KYN$ couplings comes indeed from {\sl subtracted}, 
forward dispersion relations for the spin--averaged amplitudes $D=A+\omega 
B$ {\sl only}, at variance with the $\pi N$ case, where one can use both these 
and the {\sl unsubtracted} ones for the pure $B$ amplitude as a cross--check, 
and therefore it suffers from a strong correlation to the parameters of the 
S waves at and below threshold. The simpler analytic structure of the $\pi N$ 
elastic scattering amplitudes allows even the use of {\sl partial--wave} 
dispersion relations, at variance with the $\bar KN$ case, where in some 
channels even the Born term singularities fall on the right--hand cuts$^{39}$. 
It is the very poor information on the P waves which prevents use of the $B$ 
amplitudes (dominated by these latter at low and intermediate energies): 
indeed a recent dispersive analysis has shown that even the ``best'' 
low--energy phase--shift analyses available present severe inconsistencies 
with simple tests coming from fixed--$t$ analyticity$^{39}$. New data and 
new analyses of these and of the older ones are therefore urgently needed.

\vglue 0.6truecm
\noindent{\bf 4. CAPABILITIES FOR A $\bar{K}N$--SCATTERING EXPERIMENT AT 
DA$\Phi$NE. }
\vglue 0.3truecm

DA$\Phi$NE is the $\phi$--factory (the acronym stands for ``Double Annular 
$\phi$--Factory for Nice Experiments''), which has replaced the Adone 
colliding--beam machine in the same experimental hall of the Laboratori 
Nazionali dell'I.N.F.N. in Frascati. From its expected luminosity$^{40}$ 
of $2 \times 10^{32}$ cm$^{-2}$ s$^{-1}$, and an annihilation cross section 
at the $\phi$--resonance peak of about 4.40 $\mu$b, one can see that its 
{\sl two} interaction regions will be the sources of $\simeq 436\ K^\pm$ 
s$^{-1}$, at a central momentum of 126.9 MeV/c, with the momentum 
resolution of $\simeq 1.1 \times 10^{-2}$ due to the small energy spreads 
($\Delta E/E \simeq 10^{-3}$) in the beams, as well as $\simeq 303\ K_L$ 
s$^{-1}$, at a central momentum of 110.1 MeV/c, with the slightly worse 
resolution of $\simeq 1.5 \times 10^{-2}$.

Both $\pi^\pm$'s and leptons coming out the two sources are backgrounds 
rather easy to control: the first because the $\pi^\pm$'s, though produced 
at a rate comparable to that of $K^\pm$'s (about 341 $\pi^\pm$ s$^{-1}$), 
come {\sl almost all} from events with three or more final particles, and 
can be greatly suppressed by momentum and acollinearity cuts; the second, 
as well as collinear pions from $\phi \rightarrow \pi^+ \pi^-$, produced 
at much lower rates, of order $2.5 \times 10^{-1}$ s$^{-1}$ (the leptons) 
or $3.5 \times 10^{-2}$ s$^{-1}$ (the pions), are completely eliminated 
by a momentum cut, having momenta about four times those of the $K^\pm$'s.

The interaction regions are therefore small--sized sources of 
low--momentum, tagged $K^\pm$'s and $K_L$'s, with {\sl negligible 
contaminations} (after suitable cuts on angles and momenta on the outgoing 
particles are applied {\sl event by event}), in an environment of very low 
background radioactivity: this situation is simply unattainable with 
{\sl conventional} technologies at fixed--target machines$^{41}$, 
where the impossibility of placing experiments too close to the production 
target limits from below the charged--kaon momenta, 
kaon decays in flight contaminate strongly the beams, and low--momentum 
experiments are thus possible only with the use of 
``moderators'', with a subsequent huge beam contamination at the target, as 
well as a large final--momentum spread due to straggling phenomena.

It is therefore of the highest interest to consider the feasibility of 
low--energy, $K^\pm N$ and $K_L N$ experiments at DA$\Phi$NE, with respect 
to equivalent projects at machines such as, {\sl e.g.}, KAON studied for 
TRIUMF$^{41}$ (and too hastily aborted by the Canadian government), or to 
ideas advanced for the equally sadly aborted European Hadron Factory 
project$^{42}$.

I shall, in this final part, try and give an evaluation of the rates to 
be expected in a very simple, dedicated apparatus at DA$\Phi$NE. I shall 
assume cylindrical symmetry, with a toroidal target fiducial volume, 
limited by radii $a$ and $a + d$ and of length $L$ (inside and 
outside of which you can imagine a tracking system, surrounded by a 
photon detecting system ({\sl e.g} lead--Sci--Fi sandwiches) and a 
solenoidal coil to provide the field for momentum measurements), 
filled, for simplicity, with a gas at moderate pressure. Such a detector 
immediately recalls the architecture of KLOE$^{43}$, and could be 
thought of as a much smaller (and cheaper) brother of the latter.

One must convert the usual, fixed--target expression for reaction 
rates to a spherical geometry and also include kaon decays in 
flight, getting (for simplicity this formula considers only the cases of 
either neutral kaons or zero magnetic field, but can easily be extended to 
the more general case)
$$
d N_r = [ {1 \over \rho^2} ({3\over {8\pi}})\ (L\sigma_\phi B_\phi) 
sin^2 \theta e^{-\rho / \lambda} ] \sigma_r \rho_t (\rho^2 d\rho d\Omega)\ , 
\eqno (18)
$$
with $\rho$, $\theta$ and $\phi$ spherical coordinates, $L$ the machine 
luminosity, $\sigma_\phi$ the annihilation cross section at the 
$\phi$--resonance peak, $B_\phi$ the $\phi$ branching ratio into the desired 
mode (either $K^+K^-$ or $K_LK_S$), $\sigma_r$ the reaction cross section 
for the process considered, $\rho_t$ the target {\sl nuclear} density, 
and $\lambda = p_K\tau_K / m_K$ the decay length (0.954 m for $K^\pm$'s 
and 3.429 m for $K_L$'s, at the $\phi$--resonance momenta). The small ratio 
$\lambda_+/\lambda_L$ gives immediately a reduction in radius with respect 
to KLOE of a factor from 4 to 6, larger radii for the fiducial volume 
being useless, since most of the charged kaons would have already decayed.

The reaction rate over the fiducial volume can be cast into the simple 
form (valid also in the more general case)
$$
N_r = {3\pi \over 4} r d (L \sigma_\phi B_\phi) \rho_t \sigma_r \ ,
\eqno (19)
$$
with geometrical acceptance, magnetic--field effects and kaon 
decay in flight all thrown into the reduction factor $r$, which we have 
estimated to take the  values 0.50 for $K^\pm$'s and 0.72 for $K_L$'s for 
a fiducial volume defined by $a =$ 10 cm, $d =$ 50 cm and $L =$ 1 m, to 
represent a person--sized detector, fitting in DA$\Phi$NE's second 
interaction region.

This gives, for a target volume filled by a diatomic, nearly ideal 
gas, the rates for $K^\pm$ initiated processes
$$
N_r = 10,410 \times p(\hbox{atm}) \times \sigma_r(\hbox{mb}) \ \hbox{events/y} 
\ ,
\eqno (20)
$$
for a ``physicist's year'' of $10^7$ s (for $K_L$'s the initial figure 
in the above equation is only slightly reduced by the interplay of 
$r$ and $B_\phi$ 
to 10,350), or, with rough estimates of the partial $K^- p$ cross sections 
at the $\phi$--decay momenta, to about $3 \times 10^6$ $K^-p$--initiated 
two--body events per 
year, of which about one third elastic scattering events, and the remaining 
two thirds more or less evenly divided between the five dominant inelastic 
channels ($\pi^+ \Sigma^-$, $\pi^0 \Sigma^0$, $\pi^0 \Lambda$, $\bar K^0 n$, 
and $\pi^- \Sigma^+$, more or less in order of decreasing importance). One 
could also expect about $1.5 \times 10^6$ $K_Lp$--initiated events, plus 
from 5 to 10 thousand radiative--capture events from both initial states, 
which should allow a good measurement on these processes as well$^{44}$.

These rates could be improved dramatically using liquid targets: the 
small range (1 -- 2 cm at the $\phi$--factory momenta) of kaons in liquid 
hydrogen makes the target--detector complex much smaller, but suitable only 
for measurement of inelastic or radiative--capture rates at threshold. One 
has also to weigh the reduction in cost implied by the smaller dimensions 
against the added cost of cryogeny: mentioning costs, we wish to point out 
that DA$\Phi$NE, though giving the experimenters a very small momentum range, 
could save them the cost of the tagging system needed to reject the 
contaminations of a {\sl conventional} low--energy, fixed--target 
experiment$^{45}$.

The above estimates for $K^-$ rates do not include energy losses in the 
beam--pipe wall and in the internal tracking system, which were assumed 
sufficiently thin ({\sl e. g.} of a few hundred $\mu$m of low--$Z$ 
material, such as carbon fibers or Mylar). I have indeed checked that, due 
to the shape of the angular distribution of the kaons produced, particle 
losses are rather contained and momentum losses flat around $\theta = 
\pi/2$: even for a thickness of the above--mentioned materials up to about 
1 mm, kaon momenta do not decrease appreciably below 100 MeV/c 
and losses do not grow beyond a few percents. Rather, one could exploit 
such a thickness as a ``moderator'', to span the interesting region 
of the charge--exchange threshold, measurement which would add 
additional constraints on low--energy amplitude analyses$^{1,39,46}$.

We have presented the above, oversimplified estimates to show 
that acceptable rates can be achieved, orders of magnitude above those of 
existing data at about the same momentum, {\sl i.e.} to the 
lowest--energy points of the British--Polish Track--Sensitive Target 
Collaboration, taken in the late seventies at the (R.I.P.) NIMROD 
accelerator at the then Rutherford Laboratory$^{47}$.

The statistics derived above should indeed allow a determination not only 
of the integrated cross sections for the dominant two--body channels (and, 
with a $\gamma$--detection efficiency equal to that of KLOE$^{43}$, a 
clear separation of final--state $\Sigma^0$'s from $\Lambda$'s), but also 
of those of the rarer three--body ones, plus that of the two--body angular 
distributions: the TST Collaboration$^{47}$ was able to measure $L_1$ for 
the $\pi^\pm\Sigma^\mp$ channels only, but with results consistent with 
zero within $2\sigma$, and therefore never used in the coupled--channel 
analyses. The same statistics, exploiting the self--analysing powers of 
$\Lambda$ and $\Sigma^+$ non--leptonic decays, should allow the 
determination of the polarization of the final baryons in some channels 
($K^-p\to\pi^0\Lambda, \pi^-\Sigma^+$ and $K_Lp\to\pi^+\Lambda, 
\pi^0\Sigma^+$) as well$^{46}$.

Since losses do not affect $K_L$'s, a detector of the kind 
sketched above, much smaller in size than but similar in geometry to KLOE, 
could be used without any problem to study low--energy 
$K_L \rightarrow K_S$ regeneration and charge--exchange in gaseous 
targets, providing essential information for this kind of phenomena.

I shall conclude my presentation remarking that DA$\Phi$NE (and 
$\phi$--factories in general) will present the opportunity of low--energy 
kaon experiments {\sl not feasible} (with {\sl conventional} technologies) 
at fixed--target kaon beams. Many, interesting experiments will however be 
already possible with existing detectors: KLOE$^{43}$ will surely be able to 
register all interactions of both $K^\pm$ and $K^0_L$ with the $^4$He 
filling its wire chamber, interactions never observed before at such low 
laboratory momenta, DEAR$^{48}$ will measure the K lines of kaonic hydrogen 
(and deuterium) giving independent information on the $\bar{K}N$ S--wave 
scattering lengths (and, with CCDs covering much lower $\gamma$--ray 
energies, they could also investigate the P waves through the study of 
the L lines), and FINUDA$^{49}$, though starting with a much narrower 
scope than KLOE, will anyway be able to make some high quality 
measurements, in particular of the charge--exchange processes taking 
place in the hydrogen of the plastic scintillators of its inner detector 
TOFINO.

\vglue 0.6truecm
\noindent{\bf 5. ``ENVOI''. }
\vglue 0.3truecm

We hope this last section has helped in building in the audience the 
feeling that DA$\Phi$NE is an unique opportunity, too unique to be missed, 
for bringing the quality of our information on the low--energy, $S=-1$ 
meson--baryon systems as close as possible to the one we already have on 
the $S=0$ one. To miss such an occasion would only be the sadder replay 
of another event not too far in our past, when an $e^+e^-$ machine of c.m. 
energy from 10 to 15 GeV, proposed to replace Adone (who remembers 
Super--Adone?), was killed in her crib as ``not very interesting 
physically'' (and beauty was just waiting us around the corner \dots): 
let us then hope that people and organisations footing the bills for our 
community (at least on this side of the Atlantic Ocean \dots) have learnt 
something from the misjudgements made in the past.

\vglue 1.0truecm
\leftline{\bf REFERENCES AND FOOTNOTES }
\vglue 0.3truecm

\item{1)} A.D. Martin, {\sl Nucl. Phys.} {\bf B 179} (1981) 33.

\item{2)} R. Koch, {\sl Z. Phys. C} {\bf 15} (1982) 161; G. H\"ohler, {\sl 
$\pi N$ Newslett.} {\bf 2} (1990) 1.

\item{3)} J.F. Donoghue and C.R. Nappi, {\sl Phys. Lett.} {\bf B 168} 
(1986) 105.

\item{4)} C.R. Nappi, in: {\sl ``Nuclear Chromodynamics. Quarks and Gluons 
in Particles and Nuclei''}, S.J. Brodsky and E. Moniz eds. (World 
Scientific, Singapore 1986), p. 405.

\item{5)} R.L. Jaffe and C.L. Korpa, {\sl Comments Nucl. Part. Phys.} {\bf 
17} (1987) 163; W. Kluge, in: {\sl ``Workshop on Physics and Detectors for 
{\rm DA$\Phi$NE}''}, G. Pancheri ed. (I.N.F.N., Frascati 1991), p. 469.

\item{6)} P. Amaudruz, {\sl et al.} (NMC Collaboration), {\sl Phys. Rev. 
Lett.} {\bf 66} (1991) 2712.

\item{7)} D.B. Kaplan and A. E. Nelson, {\sl Phys. Lett.} {\bf B 175} 
(1986) 57, {\sl erratum} {\bf B 179} (1986) 409; {\sl Nucl. Phys.} {\bf A 
479} (1988) 273c; A.E. Nelson and D.B. Kaplan, {\sl Phys. Lett.} {\bf B 
192} (1987) 193; {\sl Nucl. Phys.} {\bf A 479} (1988) 285c.

\item{8)} J. Gasser, {\sl Ann. Phys. (N.Y.)} {\bf 136} (1981) 62.

\item{9)} P.M. Gensini, {\sl Nuovo Cimento} {\bf 102 A} (1989) 75, {\sl 
erratum} {\bf 102 A} (1989) 1181; second order calculations in 
heavy--baryon chiral perturbation theory have been recently 
performed by B. Borasoy and U.--G. Mei\ss{ner}, {\sl Phys. Lett.} {\bf B 
365} (1996) 285, {\sl Ann. Phys. (N.Y.)} {\bf 254} (1997) 192.

\item{10)} P.M. Gensini, {\sl Perugia Univ. report}, in preparation.

\item{11)} J.M. LoSecco, in: {\sl ``Probing the Weak Interaction: CP 
Violation and Rare Decays''}, E.C. Brennan ed. (SLAC, Stanford, CA, 
1989), p. 289.

\item{12)} J. Gasser, H. Leutwyler and M.E. Sainio, {\sl Phys. Lett.} {\bf 
B 253} (1991) 260.

\item{13)} P.M. Gensini, {\sl J. Phys. G} {\bf 7} (1981) 1177.

\item{14)} P.M. Gensini, {\sl Nuovo Cimento} {\bf 103 A} (1990) 1311. 

\item{15)} P.M. Gensini, {\sl Nuovo Cimento} {\bf 103 A} (1990) 303; P.M. 
Gensini and G. Violini, {\sl $\pi N$ Newslett.} {\bf 9} (1993) 80, and {\sl 
Perugia Univ. report}, in preparation.

\item{16)} C.A. Dom\'\i nguez and E. de R\'afael, {\sl Ann. Phys. (N. Y.)} 
{\bf 174} (1987) 372; S. Narison, {\sl Riv. Nuovo Cimento} {\bf 10} (1987) 
N. 2; J.F. Donoghue, {\sl Annu. Rev. Nucl. Part. Sci.} {\bf 41} (1991) 1. For 
more recent re--determinations see: K.G. Chetyrkin, D. Pirjol and K. 
Schilcher, {\sl Phys. Lett.} {\bf B 404} (1997) 337; P. Colangelo, F. De 
Fazio, G. Nardulli and N. Paver, {\sl Phys. Lett.} {\bf B 408} (1997) 340; 
L. Lellouch, E. de R\'afael and J. Taron, {\sl Phys. Lett.} {\bf B 414} (1997) 
195, and the review by E. de R\'afael at ``QCD '97'', e-print hep-ph/9709430.

\item{17)} P.M. Gensini, {\sl Nuovo Cimento} {\bf 102 A} (1989) 1563.

\item{18)} J. Gasser, H. Leutwyler and M. E. Sainio, {\sl Phys. Lett.} {\bf 
B 253} (1991) 252; J. Gasser, {\sl $\pi N$ Newslett.} {\bf 4} (1991) 56; 
M.E. Sainio, {\sl $\pi N$ Newslett.} {\bf 4} (1991) 58, {\bf 10} (1995) 13, 
{\bf 13} (1997) 144.

\item{19)} P.M. Gensini, {\sl $\pi N$ Newslett.} {\bf 6} (1992) 21.

\item{20)} L. Maiani, G. Pancheri and N. Paver eds., {\sl ``The Second 
{\rm DA$\Phi$NE} Physics Handbook''}, (I.N.F.N., Frascati 1995), Voll. I, 
II.

\item{21)} E. Reya, {\sl Phys. Lett.} {\bf 43 B} (1973) 213; {\sl Phys. 
Rev. D} {\bf 7} (1973) 3472. 

\item{22)} B. Di Claudio, G. Violini and A.M. Rodr\'\i guez--Vargas, {\sl 
Lett. Nuovo Cimento} {\bf 26} (1979) 555; B. Di Claudio, A.M. 
Rodr\'\i guez--Vargas and G. Violini, {\sl Z. Phys. C} {\bf 3} (1979) 75; 
A.M. Rodr\'\i guez--Vargas and G. Violini, {\sl Z. Phys. C} {\bf 4} (1980) 
135; A.D. Martin and G. Violini, {\sl Lett. Nuovo Cimento} {\bf 30} (1981) 
105.

\item{23)} P.M. Gensini, {\sl Z. Phys. C} {\bf 14} (1982) 319.

\item{24)} P.M. Gensini, {\sl Nuovo Cimento} {\bf 103 A} (1990) 9.

\item{25)} P.M. Gensini, {\sl Nuovo Cimento} {\bf 84 A} (1984) 203.

\item{26)} G. Altarelli, N. Cabibbo and L. Maiani, {\sl Phys. Lett.} {\bf 
35 B} (1971) 415; {\sl Nucl. Phys.} {\bf B 34} (1971) 621.

\item{27)} M.G. Olsson and E.T. Osypowski, {\sl J. Phys. G} {\bf 6} (1980) 
423; M.G. Olsson, {\sl J. Phys. G} {\bf 6} (1980) 431.

\item{28)} A.M. Rodr\'\i guez--Vargas, {\sl Nuovo Cimento} {\bf 51 A} 
(1979) 33.

\item{29)} P.M. Gensini, {\sl Nuovo Cimento} {\bf 17 A} (1973) 557; {\sl 
Nuovo Cimento} {\bf 60 A} (1980) 221, 234, {\sl erratum} {\bf 63 A} 
(1981) 256.

\item{30)} S. Fubini and G. Furlan, {\sl Ann. Phys. (N.Y.)} {\bf 48} (1968) 
322; V. de Alfaro and C. Rossetti, {\sl Nuovo Cimento Suppl.} {\bf 6} (1968) 
575; V. de Alfaro, S. Fubini, G. Furlan and C. Rossetti, {\sl Nuovo Cimento} 
{\bf 62 A} (1969) 497.

\item{31)} C.J. Batty, {\sl Nucl. Phys.} {\bf A 411} (1983) 399; P.M. 
Gensini, {\sl Lett. Nuovo Cimento} {\bf 38} (1983) 620; P.M. Gensini, {\sl 
Nuovo Cimento} {\bf 78 A} (1983) 471.

\item{32)} T.P. Cheng and R.F. Dashen, {\sl Phys. Rev. Lett.} {\bf 26} 
(1971) 594.

\item{33)} G.D. Thompson, {\sl Lett. Nuovo Cimento} {\bf 2} (1971) 424; 
E. Reya, {\sl Phys. Rev. D} {\bf 6} (1972) 200; N.F. Nasrallah and K. 
Schilcher, {\sl Phys Rev. D} {\bf 7} (1973) 810, {\sl erratum} {\bf 15} 
(1977) 931.

\item{34)} S.A. Coon and M.D. Scadron, {\sl J. Phys. G} {\bf 18} (1992) 1923.

\item{35)} G.C. Oades, in: {\sl ``Low and Intermediate Energy Kaon-Nucleon 
Physics''}, E. Ferrari and G. Violini eds. (D. Reidel, Dordrecht 1981), 
p. 73, and private communication.

\item{36)} B. Hyams, {\sl et al.}, {\sl Nucl Phys.} {\bf B 64} (1973) 134; 
S.D. Protopopescu, {\sl et al.}, {\sl Phys. Rev. D} {\bf 7} (1973) 1279; 
C.D. Froggatt and J.L. Petersen, {\sl Nucl Phys.} {\bf B 129} (1977) 89; 
see also the review by D. Morgan and M.R. Pennington, in: {\sl ``The Second 
{\rm DA$\Phi$NE} Physics Handbook''}, L. Maiani, G. Pancheri and N. Paver 
eds. (I.N.F.N., Frascati 1995), Vol. I, p. 193.

\item{37)} R.M. Barnett, {\sl et al.} (Particle Data Group), {\sl Phys. Rev.
D} {\bf 54} (1996) 1.

\item{38)} The figures quoted come from a re--analysis of the calculations 
performed in ref. (24), due to appear shortly as a {\sl Perugia Univ. report}.

\item{39)} P.M. Gensini, R. Hurtado and G. Violini, {\sl $\pi N$ Newslett.} 
{\bf 13} 291, {\sl ibid.} 296, and e-print nucl-th/9804024, to be published in 
{\sl Genshikaku Kenkyuu} (1998); see also the University of Perugia Ph.D. 
Thesis by R. Hurtado (Perugia, March 1998), unpublished.

\item{40)} G. Vignola, in: {\sl ``Workshop on Physics and Detectors for 
{\rm DA$\Phi$NE}''}, G. Pancheri ed. (I.N.F.N., Frascati 1991), p. 11.

\item{41)} D.R. Gill ed., {\sl ``Workshop on Science at the KAON Factory''} 
(TRIUMF, Vancouver 1991).

\item{42)} J.F. Crawford ed., {\sl ``Proposal for a European Hadron 
Facility''}, {\sl report EHF 87-18} (Trieste--Mainz, May 1987).

\item{43)} J. Lee--Franzini, in: {\sl ``The Second {\rm DA$\Phi$NE} Physics 
Handbook''}, L. Maiani, G. Panche\-ri and N. Paver eds. (I.N.F.N., Frascati 
1995), Vol. II, p. 761.

\item{44)} B.P. Siegel and B. Saghai, {\sl Phys. Rev. C} {\bf 52} (1995) 392.

\item{45)} P.M. Gensini, in: {\sl ``Workshop on Physics and Detectors for 
{\rm DA$\Phi$NE}''}, G. Pancheri ed. (I.N.F.N., Frascati 1991), p. 453; T. 
Bressani, {\sl ibid.}, p. 475.

\item{46)} P.M. Gensini, in: {\sl ``The Second {\rm DA$\Phi$NE} Physics 
Handbook''}, L. Maiani, G. Pancheri and N. Paver eds. (I.N.F.N., Frascati 
1995), Vol. II, p. 739.

\item{47)} R.J. Novak, {\sl et al.}, {\sl Nucl Phys} {\bf B 139} (1978) 61; 
N.H. Bedford, {\sl et al.}, {\sl Nukleonika} {\bf 25} (1980) 509; J. 
Ciborowski, {\sl et al.}, {\sl J. Phys. G} {\bf 8} (1982) 13; D. Evans, 
{\sl et al.}, {\sl J. Phys. G} {\bf 9} (1983) 885; J. E. Conboy, {\sl et 
al.}, {\sl J. Phys. G} {\bf 12} (1986) 1143.

\item{48)} R. Baldini, {\sl et al.}, {\sl report LNF--95/055 (IR)} 
(Frascati, October 1995); C. Guaraldo, M. Braga-Direanu, M. Iliescu, 
V. Lucherini and C. Petrascu, {\sl Nucl. Phys.} {\bf A 623} 
(1997) 311c.

\item{49)} M. Agnello, {\sl et al.}, in: {\sl ``The Second {\rm 
DA$\Phi$NE} Physics Handbook''}, L. Maiani, G. Pancheri and N. Paver eds. 
(I.N.F.N., Frascati 1995), Vol. II, p. 801; A. Olin, in: {\sl ``Workshop 
on Physics and Detectors for {\rm DA$\Phi$NE} '95''}, R. Baldini, F. 
Bossi, G. Capon and G. Pancheri eds. (I.N.F.N., Frascati 1996), p. 379.

\bye